\documentclass[a4paper,12pt]{article}
\usepackage[utf8x]{inputenc}
\usepackage{amssymb}
\usepackage{amsfonts}
\usepackage{amsmath}
\usepackage{graphicx}

\def\1{\'{\i}}

\newcommand{\jmp}{J. Math. Phys.}
\newcommand{\grg}{Gen. Relativ. Gravit.}

\title{A Computer Program for the Newman-Janis Algorithm}

\author{Carlos Guti\'errez-Ch\'avez \\ 
Francisco Frutos-Alfaro \\ 
Iv\'an Cordero-Garc\1a} 
\date{\today}

\begin{document}

\maketitle

\begin{abstract}
A REDUCE code for the Newman-Janis algorithm is described.
This algorithm is intended to include rotation into nonrotating 
solutions of the Einstein field equations with spherically symmetry or
perturbed spherically symmetry and has been successfully applied to
many spacetimes. The applicability of the code is restricted to metrics
containing potentials of the form $ 1 / r $.
\end{abstract}

\section{Introduction}

\noindent
In 1965, Newman and Janis \cite{Newman-Janis65} found that it is possible, 
by means of a very peculiar complex coordinate transformation applied to 
Schwarzschild spacetime \cite{d'Inverno,Schwarzschild}, to generate the 
spinning Kerr solution \cite{d'Inverno,Kerr63} of the Einstein field equations. 
In the original paper they refer to the method as a curious {\it derivation} 
since the series of steps to obtain the desired Kerr metric do not have a 
simple or clear explanation on why they should generate a new solution 
(different from Schwarzschild) or even why those steps should provide a 
solution to vacuum field ecuations at all. However, they do mention that in a 
private communication, Kerr has shown this procedure works for the class of 
solutions $ g_{\mu \nu} = \eta_{\mu \nu} + \lambda^2 l_{\mu} l_{\nu} $  which 
contains Schwarzschild as a special case. In the same issue of the journal 
where this work was publicated, Newman {\it et al.} \cite{Newman65} used a 
similar argument and applied the exact same complex transformation to the 
Reissner-Nordstr{\"o}m metric \cite{d'Inverno,Reissner,Nordstrom} to obtain 
what they claimed to be directly shown to be a solution of the Einstein-Maxwell 
equations. Nowadays it is well known as the Kerr-Newman spinning and charged 
black hole solution. A short time after that, Demia{\'n}ski and Newman 
\cite{Demianski} succeeded in pulling out another solution by 
applying the same method to Schwarzschild metric again, but this time using a 
more involved complex coordinate transformation. The result was the Kerr-NUT 
like Demia{\'n}ski-Newman spacetime. Talbot \cite{Talbot}, on an attempt to 
explain the effectiveness of the method, briefly elaborated an argument on why 
the complex coordinate {\it trick} had been succesful on its applications so 
far. Also, he provided criteria on which metrics the proceedure should work 
according to the form of a given component of the Weyl tensor, all of this 
within the context of an application of the Newman-Penrose formalism to find 
twisting degenerate solutions to field equations. Later, Demia{\'n}ski proposed 
to find the most general solution which could be obtained by this method, 
assuming a spherical symmetric seed line element and requiring the presence of 
a non vanishing $ \Lambda $ term. He thus demonstrated that he could obtain 
the generalization of Taub-NUT including cosmological constant, but he was also 
surprised on the fact he was not able to get a version of Kerr with non 
vanishing $ \Lambda $. Although he gave an expression for his solution, it was 
later corrected by Quevedo \cite{Quevedo} who also pointed out the limitations 
of the Newman-Janis (NJ from now on) method on generating certain solutions, 
just like Demia{\'n}ski failed to obtain Kerr with a $ \Lambda $ term.

\noindent
Despite still not being fully understood and the fact that a complete 
satisfactory explanation of why it works has not been given yet, one can see
that the Newman-Janis method has proved to be successful in generating new 
stationary solutions of the Einstein field equations. Because of this 
effectiveness, applications have also been studied outside the domain of general
relativity and in various modified gravitation theories. For example, it was
shown by Krori and Bhattacharjee \cite{Krori} that the NJ technique could be 
applied within the context of Brans-Dicke theory of gravitation. Then, 
calculations were carried out to obtain not only a NUT like metric in this 
theory but also a Kerr like solution which turns out to be the rotating 
generalization of the Janis-Newman-Winicour solution \cite{JEW} for a 
spherically symmetric space time coupled to a zero rest mass scalar field.

\noindent
A REDUCE program was written, called {\texttt{Newman-Janis.red}} to this end. 
The interested reader can get our code {\texttt{Newman-Janis.red}} sending us 
an email. 

\noindent
The main goal of the code is to facilitate the application of the algorithm to 
metrics with spherically symmetry or perturbed spherically symmetry.

\section{The Newman-Janis Algorithm}

The method is easily described as a series of steps to be followed once 
one has the seed metric to which the algorithm is meant to be applied.

\begin{enumerate}
\item The seed metric in spherical coordinates needs to be transformed 
to the advanced null cordinates, also known as Eddington-Finkelstein 
coordinates \cite{d'Inverno, Drake2000, Glass2004}. 

\begin{equation*}
(t, \, r, \, \theta, \, \phi) \longrightarrow (v, \, r, \, \theta, \, \phi)
\end{equation*}

\item The next step is to find the null tetrad system that satisfies the 
contravariant metric.

\item Once the null tetrads are obtained, the Newman-Janis trick is used. The 
trick goes as follows, the radial coordinate of your metric is allowed to 
belong to a complex domain, this meaning merely that it can acquire complex 
values, but is required specifically that it must be always real, therefore 
terms of the form

\begin{equation*}
\dfrac{2}{r} \longrightarrow \dfrac{1}{r}+\dfrac{1}{{r}^{*}}
\end{equation*} 

where $ {r}^{*} $ is the complex conjugate of $ r $.

\item Then, perform the NJ complex transformation on the advanced and radial 
coordinates:

\begin{equation}
\label{one}
\begin{array}{rcl}
\bar v      & = & v + i a \cos{\theta} \\
\bar r      & = & r + i a \cos{\theta} \\
\bar \theta & = & \theta\\
\bar \phi   & = & \phi
\end{array}
\end{equation}

where $ a $ is the rotation parameter.

\item Finally, it is applied the Boyer-Lindquist coordinate transformation on 
the obtained advanced contravariant metric. 
\end{enumerate}

\noindent
In the following section, these different steps will be described as the code 
makes the calculation.

\section{The Program}

The code {\texttt Newman-Janis.red} is explained in detail as we go through the 
above exposed steps. The metric that the code needs to run, has the form:

\begin{equation}
\label{two}
ds^{2}=g_{tt} \, {d t}^{2} - g_{rr} \, {d r}^{2} - g_{th} \, {d \theta}^{2}
- g_{pp} \, {d \phi}^{2} ,
\end{equation}

\noindent
where one has to define explicitly the metric components 
($ g_{tt}, \, g_{rr}, \, g_{th}, \, g_{pp} $). A subroutine finds the generalized 
Eddington-Finkelstein transformation: 

\begin{equation}
\label{three}
d v = d t + \sqrt{\frac{g_{rr}}{g_{tt}}} \, d r 
\end{equation}

\noindent
in terms of the seed metric. 

\noindent
Now, we have the metric in terms of the advanced null coordinates: 

\begin{equation}
\label{four}
ds^2 = g_{tt} \, {d v}^2 + 2 g_{vr} \, d v \, d r - g_{th} d \theta^2 
- g_{pp} d \phi^2 ,
\end{equation}

\noindent
where

$$ g_{vr} = - \sqrt{g_{tt} \, g_{rr}} . $$

\noindent
The code enlists the components of this new metric tensor, writes it in matrix 
notation to calculate the inverse matrix. Then, the null tetrads are computed 
in terms of the components of this metric. To avoid errors the program computes 
the contravariant metric and verifies that both the tetrads and the metric 
components fulfill the following relation

\begin{equation}
\label{five}
g^{ij}=l^{i}n^{j}+n^{i}l^{j}-m^{i}\bar{m}^{j}-\bar{m}^{i}m^{j}
\end{equation}

\noindent 
The step 3 of the Newman-Janis procedure can only be done by hand, this is 
because it is cumbersome to do it with REDUCE, and it maybe impossible to 
compute at all.     

\noindent 
The next step is to apply the Newman-Janis transformation (\ref{one}) to the 
latter obtained null tetrads, which is the key step in the whole process. For 
the sake of simplicity in notation the code displays the following quantity in 
all the tetrads expressions 

\begin{equation}
\label{six}
\rho^{2}=r^{2}+(a \cos\theta)^{2}
\end{equation}

\noindent 
Then, the {\it new} contravariant metric components is obtained using equation 
(\ref{five}). The expression for it is of the form

\begin{equation}
\label{seven}
d s^2 = g_{tt} \, {d t}^2 + 2 g_{vr} \, dv \, dr + 2 g_{vp} \, d v \, d \phi 
+ 2 g_{rp} \, d r \, d \phi + g_{th} \, d \theta^2 + g_{pp} \, d \phi^2 .
\end{equation}

\noindent 
The {\it new} covariant metric is determined from the 
contravariant one. The code computes again the new covariant metric in a more 
compacted way and confirmes that both expressions are equivalent by performing 
the difference between them.

\noindent
Next, the transformation to the generalized Boyer-Lindquist coordinates is 
performed by the program in order to display the final metric in the standard 
form. The code rewrites the expressions in a simpler and standard way: 

\begin{equation}
\label{eight}
d s^2 = g_{tt} \, {d t}^2 - (\alpha \, g_{vr} + \beta \, g_{rp}) \, {d r}^2 
+ g_{pp} \, d \phi^2 + 2 g_{vp} \, d t \, d \phi + g_{th} \, d \theta^2
\end{equation}

\noindent
and compares them to avoid mistakes. In (\ref{eight}) we have used  

\begin{equation*}
\begin{array}{rcl}
\alpha & = & (g_{pp} \, g_{vr} - g_{rp} \, g_{vp}) / \gamma , \\
\beta  & = & (g_{rp} \, g_{tt} - g_{vp} \, g_{vr}) / \gamma , \\
\gamma & = &  g_{tt} \, g_{pp} - g_{vp}^2 . 
\end{array}
\end{equation*}

\section{Test Results}

\noindent
We tested the program for the Schwarzschild \cite{Schwarzschild} and 
Brans-Dicke metrics \cite{Krori}. The first metric is given by  

\begin{equation*}
{d} s^2 = \left(1 - \frac{R_s}{r} \right) \, d t^2
- \left(1 - \frac{R_s}{r} \right)^{-1} \, d r^2 - r^2 \, d \Omega^2 ,
\end{equation*}

\noindent
where $ R_s = 2 M $, and 
$ {d} \Omega^2 = d \theta^2 + \sin^2{\theta} \, d \phi^2 . $ The output was 
the Kerr metric as expected \cite{d'Inverno,Kerr63}:

\begin{equation*}
{d} {s}^{2} = \frac{\Delta}{{\rho}^2} [d t - a {\sin}^2 {\theta} d \phi]^2 
- \frac{{\sin}^2 {\theta}}{{\rho}^2} [(r^2 + a^2) d \phi - a d t]^2 
- \frac{{\rho}^2}{\Delta} d {r}^2 - {\rho}^2 d {\theta}^2 . 
\end{equation*}

\noindent
where $ \Delta = r^2 - R_s r + a^2 $.

\noindent
The second metric is given by  

\begin{equation*}
{d} s^2 = \left(1 - \frac{R_s}{r} \right)^{\eta} \, d t^2
- \left(1 - \frac{R_s}{r} \right)^{\xi-1} \, d r^2 
- r^2 \left(1 - \frac{R_s}{r} \right)^{\xi} \, d \Omega^2 , 
\end{equation*}

\noindent
where $ \eta $, and $ \xi $ are constant.

\noindent
The resulting metric \cite{Krori}  is given by

\begin{eqnarray*}
{d} s^2 & =  & \left(1 - \frac{R_s r}{\rho^2} \right)^{\eta} \, 
[d t + \omega \, d \phi]^2 \\
& - & \rho^2 \left(1 - \frac{R_s r}{\rho^2} \right)^{\xi} \, 
\left(\frac{d r^2}{\Delta} + d \Omega^2 \right) \nonumber \\
& - & 2 \left(1 - \frac{R_s r}{\rho^2} \right)^{\sigma} \, 
[d t + \omega \, d \phi] d \phi , \nonumber
\end{eqnarray*}

\noindent
where $ \omega = a \sin^2{\theta} $.

\section{Conclusions}

\noindent
This REDUCE program is very useful to include rotation to metrics with 
spherical symmetry. It should not be used for metrics with cosmological 
constant and with no spherical symmetry. The inputs to the program, the user 
has to provide, are the metric and the change in term like 
$ {2}/{r} \longrightarrow {1}/{r}+{1}/{{r}^{*}} $. Moreover, it was 
successfully tested with the Schwarzschild and the Brans-Dicke metrics.
At the moment, there is no standard procedure to include rotation into metrics 
with no other than spherical symmetry, but if in the future it could be 
possible, then this code can be an initial step towards other programs to 
attack that problem.

\section*{Appendix}

\noindent
The output for main result in the case of the Schwarzschild metric is:

\smallskip
\noindent
\texttt{g(0,0) := coeffn(ds2,dt,2); \\ g(0,0) := ( - 2*m*r + rho**2)/rho**2}

\noindent
\texttt{g(0,3) := coeffn(ds2,dt,1)/(2*dphip); \\ 
g(0,3) := ( - 2*sin(theta)**2*a*m*r)/rho**2}

\noindent
\texttt{g(3,0) := coeffn(ds2,dphip,1)/(2*dt); \\
g(3,0) := ( - 2*sin(theta)**2*a*m*r)/rho**2}

\noindent
\texttt{g(1,1) := coeffn(ds2,dr,2); \\
g(1,1) := ( - rho**2)/(sin(theta)**2*a**2 - 2*m*r + rho**2)}

\noindent
\texttt{g(2,2) := coeffn(ds2,dtheta,2); \\ g(2,2) := - rho**2}

\noindent
\texttt{g(3,3) := coeffn(ds2,dphip,2); \\
g(3,3) := (sin(theta)**2*( - 2*sin(theta)**2*a**2*m*r \\
- sin(theta)**2*a**2*rho**2 - rho**4))/rho**2}

\bigskip
\noindent
The output for main result in the case of the Brans-Dicke metric is:

\smallskip
\noindent
\texttt{g(0,0) := coeffn(ds2,dt,2); \\
g(0,0) := (( - 2*r*rs + rho**2)/rho**2)**eta}

\noindent
\texttt{g(0,3) := coeffn(ds2,dt,1)/(2*dphip); \\
g(0,3) := sin(theta)**2*a*( \\
- sqrt(( - (( - 2*r*rs + rho**2)/rho**2)**(chi + eta))/(2*r*rs \\ 
- rho**2))*rho + (( - 2*r*rs + rho**2)/rho**2)**eta)}

\noindent
\texttt{g(3,0) := coeffn(ds2,dphip,1)/(2*dt); \\
g(3,0) := sin(theta)**2*a*( \\ 
- sqrt(( - (( - 2*r*rs + rho**2)/rho**2)**(chi + eta))/(2*r*rs \\ 
- rho**2))*rho + (( - 2*r*rs + rho**2)/rho**2)**eta)}

\noindent
\texttt{g(1,1) := coeffn(ds2,dr,2); \\
g(1,1) := ( - (( - 2*r*rs \\ 
+ rho**2)/rho**2)**chi*rho**2)/(sin(theta)**2*a**2 - 2*r*rs \\ 
+ rho**2)}

\noindent
\texttt{g(2,2) := coeffn(ds2,dtheta,2); \\ 
g(2,2) := - (( - 2*r*rs + rho**2)/rho**2)**chi*rho**2}

\noindent
\texttt{g(3,3) := coeffn(ds2,dphip,2); \\
g(3,3) := sin(theta)**2*( - 2*sqrt(( \\ 
- (( - 2*r*rs + rho**2)/rho**2)**(chi + eta))/(2*r*rs \\
- rho**2))*sin(theta)**2*a**2*rho \\ 
- (( - 2*r*rs + rho**2)/rho**2)**chi*rho**2 \\
+ (( - 2*r*rs + rho**2)/rho**2)**eta*sin(theta)**2*a**2)}


\begin{thebibliography}{99}

\bibitem{Boyer1967}
R.~H.~Boyer, R.~W.~Lindquist,
\newblock Maximal analytic extension of the Kerr metric.
\newblock {\em \jmp}, 8(2):265--281, 1967.

\bibitem{Demianski}
M.~Demia{\'n}ski, E.~T.~Newman, 
\newblock Combined Kerr-NUT Solution of the Einstein Field Equations.
\newblock Bulletin de l’Acad\'emie Polonaise des Sciences, 14, 653--657, 1966.

\bibitem{d'Inverno}
R.~d'Inverno,
\newblock {\em Introducing Einstein's Relativity}.
\newblock Oxford University Press, USA, 1992.

\bibitem{Drake2000}
S.~P.~Drake and P.~Szekeres,
\newblock Uniqueness of the Newman-Janis algorithm in generating the 
kerr-newman metric.
\newblock {\em \grg}, 32:445--458, Mar 2000.

\bibitem{Flaherty1976}
E.~J.~Flaherty,
\newblock {\em Hermitian and K\"ahlerian Geometry in Relativity}
(Lecture Notes in Physics 46).
\newblock Springer-Verlag, 1976.

\bibitem{Glass2004}
E.~N.~Glass, J.~P.~Krisch,
\newblock Spinning up asymptotically flat spacetimes.
\newblock {\em Classical and Quantum Gravity}, 21(23):5543, 2004.

\bibitem{Hearn1999}
A.~C.~Hearn, 
\newblock {\em REDUCE} (User's and Contributed Packages Manual). 
\newblock Konrad-Zuse-Zentrum f\"ur Informationstechnik, Berlin, 1999.

\bibitem{Ibohal2005}
N.~Ibohal,
\newblock Rotating metrics admitting non-perfect fluids.
\newblock {\em \grg}, 37:19--51, 2005.
\newblock 10.1007/s10714-005-0002-6.

\bibitem{JEW}
A.~I.~Janis, E.~T.~Newman, and J.~Winicour,
\newblock Reality of the Schwarzschild Singularity.
\newblock {\em Phys. Rev.}, {\bf 20} (16), 878--880, 1968.

\bibitem{Kerr63}
R.~P.~Kerr.
\newblock Gravitational field of a spinning mass as an example of algebraically 
special metrics.
\newblock {\em Phys. Rev. Lett.}, 11:237--238, Sep 1963.

\bibitem{Kerr-Schild}
R.~P.~Kerr, A.~Schild,  
\newblock Some algebraically degenerate solutions of Einstein's gravitational 
field equations.
\newblock {\em Applications of Nonlinear Partial Differential Equations in 
Mathematical Physics} (R.~Finn ed.)
\newblock Proceedings of Symposia in Applied Mathematics 
(American Mathematical Society, Providence, Rhode Island), Vol. XVII, 199--209, 
1965.

\bibitem{Hongsu99}
H.~Kim,
\newblock Spinning BTZ black hole versus Kerr black hole: A closer look.
\newblock {\em Phys. Rev. D}, 59:064002, Feb 1999.

\bibitem{Krori}
K.~D.~Krori, D.~R.~Bhattacharjee, 
\newblock Kerr-like metric in Brans-Dicke theory
\newblock Journal of Mathematical Physics, 23 (4), 637--638, 1982.

\bibitem{Newman-Janis65}
E.~T.~Newman and A.~I.~Janis,
\newblock Note on the Kerr spinning-particle metric.
\newblock {\em \jmp}, 6(6):915--917, 1965.

\bibitem{Newman65}
E.~T.~Newman, E.~Couch, K.~Chinnapared, A.~Exton, A.~Prakash, R.~Torrence,
\newblock Metric of a rotating, charged mass.
\newblock {\em \jmp}, 6(6):918--919, 1965.

\bibitem{Nordstrom}
G.~Nordstr{\o}m,
\newblock On the Energy of the Gravitational Field in Einstein’s Theory 
\newblock {\em Verhandl. Koninkl. Ned. Akad. Wetenschap.}, {\bf 20}, 
1238--1245, 1918.

\bibitem{Quevedo}
H.~Quevedo,
\newblock Complex Transformations of the Curvature Tensor.
\newblock General Relativity and Gravitation, 24 (7), 693--703, 1992

\bibitem{Reissner}
H.~Reissner. 
\newblock \"Uber die Eigengravitation des elektrischen Feldes nach der
Einsteinschen Theorie. 
\newblock {\em Annalen der Physik}, {\bf 50}, 106--120, 1916.

\bibitem{Schwarzschild}
K.~Schwarzschild,  
\newblock \"Uber das Gravitationsfeld eines Massen\-punk\-tes nach der 
Ein\-stein\-schen Theorie. 
\newblock {\em Sitzungsberichte der Königlich Preussischen Akademie der 
Wissenschaften}, {\bf 7}, 189--196. 1916 (ArXiv:physics/9905030).

\bibitem{Talbot}
C.~J.~Talbot, 
\newblock Newman-Penrose Approach to Twisting Degenerate Metrics.
\newblock Communications in Mathematical Physics, 13, 45--61, 1969.

\end{thebibliography}
\end{document}